\documentclass[draftcls]{IEEEtran}
\usepackage{epsfig,makeidx,color}
\usepackage{graphicx}
\usepackage{amsbsy}
\usepackage{amsmath}
\usepackage{amssymb}
\usepackage{euscript}

\newtheorem{problem}{\bf Problem}

\newcommand{\ave}{{\mathbb E}}

\newcommand {\Define} {\stackrel {\Delta} {=}  }

\begin{document}

\title{Precoding by Pairing Subchannels to Increase\\ MIMO Capacity with Discrete Input Alphabets}
\author{Saif~Khan~Mohammed,~\IEEEmembership{Student~Member,~IEEE,}
        Emanuele~Viterbo,~\IEEEmembership{Senior~Member,~IEEE}
        Yi~Hong,~\IEEEmembership{Member,~IEEE,}
        and~Ananthanarayanan Chockalingam,~\IEEEmembership{Senior~Member,~IEEE}
\thanks{\scriptsize  S. K. Mohammed and A. Chockalingam are with Indian Institute of Science,
Bangalore $560012$, India. E-mail: $\tt saifind2007$@$\tt yahoo.com$ and $\tt achockal$@$\tt ece.iisc.ernet.in$.
Saif K. Mohammed is currently visiting DEIS, Universit\`a della Calabria, Italy.}
\thanks{\scriptsize Yi Hong and Emanuele Viterbo are with DEIS -
Universit\`{a} della Calabria, via P. Bucci, 42/C, 87036 Rende (CS),
Italy.  E-mail: $\tt \{hong,viterbo\}$@$\tt deis.unical.it$.}}

\onecolumn
\maketitle

\begin{abstract}
We consider Gaussian multiple-input multiple-output (MIMO) channels with discrete input alphabets.
We propose a non-diagonal precoder based on the X-Codes in \cite{Xcodes_paper} to increase the mutual information.
The MIMO channel is transformed into a set of parallel subchannels using
Singular Value Decomposition (SVD) and X-Codes are then used to pair the subchannels.
X-Codes are fully characterized by the pairings and a $2\times 2$ real rotation matrix for each pair (parameterized with a single angle).
This precoding structure enables us to express the total mutual information as a sum
of the mutual information of all the pairs.
The problem of finding the optimal precoder with the above structure,
which maximizes the total mutual information,
is solved by  {\em i}) optimizing the rotation angle and the power allocation within each pair and
{\em ii}) finding the optimal pairing and power allocation among the pairs.
It is shown that the mutual information achieved with the proposed pairing scheme
is very close to that achieved with the optimal precoder by Cruz {\em et al.}, and is significantly
better than Mercury/waterfilling strategy by Lozano {\em et al.}.
Our approach greatly simplifies both the precoder optimization and the detection complexity,
making it suitable for practical applications.
\end{abstract}


\IEEEpeerreviewmaketitle
%%%%%%%%%%%%%%%%%%%%%%%%%%%%%%%%%%%%%%%%%%%%%%%%%%%%%%%%%%%%%%%%%%%%%%%%%%%%%%%%%%%%%%%%%%%%%%%%%
\section{Introduction}
%%%%%%%%%%%%%%%%%%%%%%%%%%%%%%%%%%%%%%%%%%%%%%%%%%%%%%%%%%%%%%%%%%%%%%%%%%%%%%%%%%%%%%%%%%%%%%%%%
% no \IEEEPARstart
Many modern communication channels are modeled as a Gaussian multiple-input multiple-output (MIMO) channel.
Examples include multi-tone digital subscriber line (DSL), orthogonal frequency division multiplexing (OFDM)
and multiple transmit-receive antenna systems.
%% with channel state information at transmitter (CSIT).
It is known that the capacity of the Gaussian MIMO channel is achieved by beamforming a {\em  Gaussian input alphabet} along the
right singular vectors of the MIMO channel.  The received vector is projected along the left singular vectors, resulting in a set of parallel Gaussian subchannels.
Optimal power allocation between the subchannels is achieved by waterfilling \cite{Cover}.
In practice, the input alphabet is {\em not Gaussian} and is generally chosen from a finite signal set.

We distinguish between two kinds of MIMO channels: {\em i}) {\em diagonal} (or parallel) channels and
{\em ii})  {\em non-diagonal} channels.

For a diagonal MIMO channel with discrete input alphabets, assuming only power allocation on each subchannel (i.e., a diagonal precoder),
Mercury/waterfilling was shown to be optimal by Lozano {\em et al.} in \cite{Lozano}.
With discrete input alphabets,
Cruz {\em et al.} later proved in  \cite{cruz} that the optimal precoder is,
however, non-diagonal, i.e., precoding needs to be performed across all the subchannels.

For a general non-diagonal Gaussian MIMO channel, it was also shown in \cite{cruz}
that the optimal precoder is non-diagonal.
Such an optimal precoder is given by a fixed point equation, which requires a high
complexity numeric evaluation.
Since the precoder jointly codes all the $n$ inputs, joint decoding is also required at the receiver.
Thus, the decoding complexity can be very high, specially for large $n$,
as in the case of DSL and OFDM applications.
This motivates our quest for a practical low complexity precoding scheme achieving near optimal capacity.

In this paper, we consider a general MIMO channel and a non-diagonal precoder based on X-Codes \cite{Xcodes_paper}.
The MIMO channel is transformed into a set of parallel subchannels
using Singular Value Decomposition (SVD) and
X-Codes are then used to pair the subchannels.
X-Codes are fully characterized by the pairings and the 2-dimensional real rotation matrices for each pair. These rotation matrices are parameterized with a single angle.
This precoding structure enables us to express the total mutual information as a sum of the mutual information of all the pairs.
%Hence, the power allocation can be split into {\em i})
%power allocation among the pairs and {\em ii}) power allocation within each pair.

The problem of finding the optimal precoder with the above structure,
which maximizes the total mutual information,
can be split into two  tractable problems:
{\em i}) optimizing the rotation angle and the power allocation within each pair and
{\em ii}) finding the optimal pairing and power allocation among the pairs.
It is shown by simulation that the mutual information achieved with the proposed pairing scheme
is very close to that achieved with the optimal precoder in \cite{cruz}, and is significantly
better than the Mercury/waterfilling strategy in \cite{Lozano}.
Our approach greatly simplifies both the precoder optimization and the detection complexity,
making it suitable for practical applications.

The rest of the paper is organized as follows.
Section~\ref{SMPrecoding} introduces the system model and SVD precoding.
In Section~\ref{optimalprecoding}, we provide a brief review of the optimal precoding with
discrete inputs in \cite{cruz} and the relevant MIMO capacity.
In Section~\ref{PrecodingX}, we present the precoding using X-Codes with discrete
inputs and the relevant capacity expressions.
In Section~\ref{two_subch}, we consider the first problem, which is to find the
optimal rotation angle and power allocation for a given pair.
This problem is equivalent to optimizing the mutual information for a Gaussian
MIMO channel with two subchannels.
In Section \ref{multi_subch}, using the results from Section \ref{two_subch},
we attempt to optimize the mutual information for a Gaussian MIMO channel
with $n$ subchannels, where $n>2$.
Conclusions are drawn in Section~\ref{conclusions}.Finally, in Section \ref{sec_ofdm} we discuss the application of our precoding to OFDM systems.

{\em Notations}:
The field of complex numbers is denoted by $\mathbb{C}$  and let ${\mathbb R}^+$ be the positive real numbers.
Superscripts $^T$ and $^{\dag}$ denote transposition and Hermitian transposition, respectively.
The $n\times n$ identity matrix is denoted by $\mathbf{I}_{n}$, and the zero matrix is denoted by
$\mathbf{0}$.
The $\ave[\cdot]$ is the expectation operator, $\Vert \cdot \Vert$ denotes the
Euclidean norm of a vector, and $\|\cdot\|_F$ the Frobenius norm of a matrix.
%Furthermore, $\lfloor c \rfloor$ denotes the largest integer less than $c$. We let $\Re(\cdot)$ and $\Im(\cdot)$
%denote the real and imaginary parts of a complex argument.
Finally, we let $\mbox{tr}(\cdot)$ be the trace of a matrix.

%%%%%%%%%%%%%%%%%%%%%%%%%%%%%%%%%%%%%%%%%%%%%%%%
\section{System model and Precoding with Gaussian inputs}\label{SMPrecoding}
%%%%%%%%%%%%%%%%%%%%%%%%%%%%%%%%%%%%%%%%%%%%%%%%
%In this section, we present the Gaussian MIMO system model and define its capacity.
We consider a $n_t\times n_r$ MIMO channel,
where the channel state information (CSI) is known perfectly at both transmitter and receiver.
Let ${\bf x} = (x_1,\cdots, x_{n_t})^T$ be the vector of input symbols to the channel,
and let ${\bf H}=\{h_{ij}\}$, $i=1, \cdots, n_r$, $j=1, \cdots, n_t$, be a full rank $n_r\times n_t$
channel coefficient matrix, with $h_{ij}$ representing the complex channel gain between the $j$-th input symbol
and the $i$-th output symbol.
The vector of $n_r$ channel output symbols is given by
\begin{equation}
\label{system_modeleq}
{\bf y} = \sqrt {P_T}{\bf H}{\bf x} + {\bf w}
\end{equation}
where ${\bf w}$ is an uncorrelated Gaussian noise vector, such that
$\ave[{\bf w}{\bf w}^\dag]= {\bf I}_{n_r}$, and
$P_T$ is the total transmitted power.
The power constraint is given by
\begin{equation}
\label{tx_pow}
\ave[\Vert {\bf x} \Vert^2] = 1
\end{equation}
The maximum multiplexing gain of this channel is $n = \min(n_r,n_t)$.
Let ${\bf u}=(u_1,\cdots, u_{n})^T \in{\mathbb C}^{n}$ be the vector of $n$ information symbols
to be sent through the MIMO channel, with $\ave[\vert u_i \vert^2] = 1, i = 1, \cdots, n$.
Then the vector ${\bf u}$ can be precoded using a $n_t \times n$ matrix ${\bf T}$,
resulting in ${\bf x}={\bf T}{\bf u}$.

The capacity of the deterministic Gaussian MIMO channel is then achieved by solving
\begin{problem}\label{cap_gaussian}
\begin{eqnarray}
\label{cap_gaussian_mimo}
C({\bf H},P_T) & =  & \max_{ {\bf K}_{\bf x} | \mbox{tr}({\bf K}_{\bf x} ) = 1} I({\bf x} ; {\bf y} | {\bf H}) \\ \nonumber
      & \geq & \max_{ {\bf K}_{\bf u}, {\bf T}  \,| \,
      \mbox{tr}({\bf T}{\bf K}_{\bf u}{\bf T}^\dag)  = 1} I({\bf u} ; {\bf y} | {\bf H})
\end{eqnarray}
\end{problem}
where $I({\bf x} ; {\bf y} | {\bf H})$ is the mutual information between ${\bf x}$
and ${\bf y}$, and ${\bf K}_{\bf x} \Define \ave[{\bf x}{\bf x}^\dag]$,
${\bf K}_{\bf u} \Define \ave[{\bf u}{\bf u}^\dag]$ are the covariance matrices of ${\bf x}$ and ${\bf u}$ respectively.
The inequality in (\ref{cap_gaussian_mimo}) follows from the data processing inequality \cite{Cover}.

Let us consider the singular value decomposition (SVD) of the channel
${\bf H}={\bf U}{\mathbf \Lambda}{\bf V}$, where ${\bf U} \in {\mathbb C}^{n_r \times n}$,
${\mathbf \Lambda} \in {\mathbb C}^{n \times n}$, ${\bf V} \in {\mathbb C}^{n \times n_t}$, ${\bf U}^\dag{\bf U}={\bf V}{\bf V}^\dag={\bf I}_{n}$,
and ${\mathbf \Lambda} = \mbox{diag} (\lambda_{1}, \ldots, \lambda_{n})$
with $\lambda_1   \geq \lambda_2, \cdots, \geq \lambda_{n} \geq 0$.

Telatar showed in  \cite{tela99} that the Gaussian MIMO capacity $C({\bf H},P_T)$, is achieved when
${\bf x}$ is Gaussian distributed and ${\bf T}{\bf K}_{\bf x} {\bf T}^{\dag}$ is diagonal.
Diagonal ${\bf T}{\bf K}_{\bf x} {\bf T}^{\dag}$  can be achieved by using the optimal precoder
matrix ${\bf T} =  {\bf V}^\dag {\bf P}$,
where ${\bf P} \in ({\mathbb R^{+}})^{n}$ is the diagonal power allocation matrix
such that $\mbox{tr}( {\bf P}{\bf P}^\dag) = 1$.
Furthermore, $u_i, i = 1, \ldots, n$, are i.i.d. Gaussian (i.e., {\em no coding is required across the input symbols $u_i$}).
With this, the second line of (\ref{cap_gaussian_mimo}) is actually an equality.
Also, projecting the received vector ${\bf y}$ along the columns
of ${\bf U}$ is information lossless and transforms the non-diagonal MIMO channel into an equivalent diagonal channel with
$n$ non-interfering subchannels. The equivalent diagonal system model is then given
by
\begin{equation}
\label{eq_diag_model_ch6}
{\bf r} \Define {\bf U}^{\dag}{\bf y} = \sqrt{P_T} {\mathbf \Lambda} {\bf P} {\bf u}  + {\Tilde {\bf w}}
\end{equation}
where ${\Tilde {\bf w}}$ is the equivalent noise vector, and has the same statistics as ${\bf w}$.
The total mutual information is now given by
\begin{equation}
\label{tot_mi_gauss_ch6}
I({\bf x} ; {\bf y} | {\bf H})  = \sum_{i=1}^{n} \log_2(1 + {\lambda_i}^2 p_{i}^2 P_T)
\end{equation}
Note that now the mutual information is a function of only the
power allocation matrix ${\bf P}=$ diag$(p_1,\ldots,p_n)$, with the constraint
tr$({\bf P}{\bf P}^{\dag}) = 1$.
Optimal power allocation is achieved through waterfilling between the $n$ parallel channels of the equivalent system in (\ref{eq_diag_model_ch6}) \cite{Cover}.

%%%%%%%%%%%%%%%%%%%%%%%%%%%%%%%%%%%%%%%%%%%%%%%%%%%%%%%%%%
\section {Optimal precoding with discrete inputs}\label{optimalprecoding}
%%%%%%%%%%%%%%%%%%%%%%%%%%%%%%%%%%%%%%%%%%%%%%%%%%%%%%%%%%
In practice, discrete input alphabets are used. Subsequently, we assume that the $i$-th
information symbol is given by $u_i \in {\mathcal U}_i$, where ${\mathcal U}_i \subset {\mathbb C}$
is a finite signal set.
Let ${\mathcal S} \Define {\mathcal U}_1 \times {\mathcal U}_2 \times  \cdots \times {\mathcal U}_n$ be the overall input alphabet.
The capacity of the Gaussian MIMO channel with discrete input alphabet
${\mathcal S}$ is defined by the following problem
\begin{problem}\label{cap_discrete}
\begin{eqnarray}
\label{cap_gaussian_mimo2}
C_{\mathcal S}({\bf H},P_T) =  \max_{{\bf T} \,| \,
   {\bf u} \in {\mathcal S}, \Vert {\bf T} \Vert_F = 1} I({\bf u} ; {\bf y} | {\bf H})
\end{eqnarray}
\end{problem}
Note that there is no maximization over the pdf of ${\bf u}$, since we fix ${\bf K}_{\bf u} = {\bf I}_n$.
The optimal precoder ${\bf T}^{*}$, which solves Problem \ref{cap_discrete}, is given by the following fixed point equation given in \cite{cruz}
\begin{equation}
\label{opt_precoder}
{\bf T}^{*} = \frac {{\bf H}^\dag{\bf H}{\bf T}^{*} {\bf E} } {\Vert  {\bf H}^\dag{\bf H}{\bf T}^{*} {\bf E}   \Vert_F}
\end{equation}
where ${\bf E}$ is the minimum mean-square error (MMSE) matrix of ${\bf u}$ given by
\begin{equation}
\label{mmse_mat}
{\bf E} = \ave [ ( {\bf u} - \ave[{\bf u}|{\bf y}])  ( {\bf u} - \ave[{\bf u}|{\bf y}])^\dag  ]
\end{equation}
The optimal precoder is derived using the relation between MMSE and mutual information \cite{guo_mmse}.
We observe that, with discrete input alphabets, it is no longer optimal to beamform along the column
vectors of ${\bf V}^\dag$ and then use waterfilling on the parallel subchannels.
Even when ${\bf H}$ is diagonal (parallel non-interfering subchannels), the optimal precoder
${\bf T}^{*}$ is {\em non diagonal}, and can be computed numerically (using a gradient based method)
as discussed in \cite{cruz}.
However, the complexity of computing ${\bf T}^{*}$ is prohibitively high for practical
applications, especially when $n$ is large and/or the channel changes frequently.
%This makes, achieving the optimal capacity impractical.

We propose a suboptimal precoding scheme based on X-Codes \cite{Xcodes_paper},
which achieves close to the optimal capacity $C_{\mathcal S}({\bf H},P_T)$,
at low encoding and decoding complexities.

%%%%%%%%%%%%%%%%%%%%%%%%%%%%%%%%%%%%%%%%%%%%%%%%%%%%%%%%%%
\section {Precoding with X-Codes}\label{PrecodingX}
%%%%%%%%%%%%%%%%%%%%%%%%%%%%%%%%%%%%%%%%%%%%%%%%%%%%%%%%%%
X-Codes are based on a pairing of $n$ subchannels
$\ell = \{ (i_k,j_k)\in [1,n]\times [1,n], i_k < j_k, k = 1, \ldots n/2 \}$.
For a given $n$, there are $(n-1)(n-3) \cdots 3\,1$ possible pairings.
Let ${\mathcal L}$ denote the set of all possible pairings.
For example, with $n = 4$, we have
\[
{\mathcal L} = \left\{  \{(1,4),(2,3)\}  \,,\, \{(1,2),(3,4)\} \,,\,  \{(1,3),(2,4)\}  \right \}
\]

X-Codes are generated by a $n \times n$ real orthogonal matrix, denoted by ${\bf G}$.
When precoding with X-Codes, the precoder matrix is given by ${\bf T} =  {\bf V}^\dag {\bf P}{\bf G}$,
where ${\bf P} = \mbox{diag}(p_1, p_2, \cdots, p_n) \in {(\mathbb R^{+})}^{n}$ is the diagonal power
allocation matrix such that $\mbox{tr}( {\bf P}{\bf P}^\dag) = 1$.
The $k$-th pair consists of subchannels $i_k$ and $j_k$. For
the $k$-th pair, the information symbols $u_{i_k}$ and $u_{j_k}$
are jointly coded using a $2 \times 2$ real orthogonal matrix ${\bf A}_k$ given by
\begin{equation}
\label{akmat}
{\bf A}_k = \left[\begin{array}{cc}
\cos(\theta_k) & \sin(\theta_k) \\
-\sin(\theta_k) &  \cos(\theta_k)
\end{array} \right] \ \ \ k=1,\ldots n/2
\end{equation}
The angle $\theta_k$ can be chosen to maximize the mutual information for the $k$-th pair.
Each ${\bf A}_k$ is a submatrix of the code matrix ${\bf G}=(g_{i,j})$ as shown below
\begin{eqnarray}\label{Xak}
\begin{array}{ll}
g_{{i_k},{i_k}} =  \cos(\theta_k) & g_{{i_k},{j_k}} = \sin(\theta_k) \\
g_{{j_k},{i_k}} = -\sin(\theta_k) & g_{{j_k},{j_k}} = \cos(\theta_k)
\end{array}
\end{eqnarray}
%where $g_{i,j}$ is the entry of ${\bf G}$ in the $i$-th row and $j$-th column.
It was shown in \cite{Xcodes_paper} that, for achieving the best diversity gain, an optimal
pairing is one in which the $k$-th subchannel is paired with the $(n - k + 1)$-th subchannel.
For example, with this pairing and $n$ = $6$, the X-Code generator matrix is given by
\[ \footnotesize {\mathbf G}\!\!=\!\!\left[ \!\!\begin{array}{cccccc}
\cos(\theta_1) & ~~ & ~~ & ~~ & ~~ & \sin(\theta_1) \\
~~ & \cos(\theta_2) & ~~ & ~~ & \sin(\theta_2) & ~~ \\
~~ & ~~ & \cos(\theta_3) & \sin(\theta_3) & ~~ & ~~ \\
~~ & ~~ & -\sin(\theta_3) & \cos(\theta_3) & ~~ & ~~ \\
~~ & -\sin(\theta_2) & ~~ & ~~ & \cos(\theta_2) & ~~ \\
-\sin(\theta_1) & ~~ & ~~ & ~~ & ~~ & \cos(\theta_1)
\end{array} \!\! \right]\]
The special case with $\theta_k = 0, k = 1,2, \cdots, n/2$, results in no coding across subchannels.

Given the generator matrix ${\bf G}$, the subchannel gains ${\mathbf \Lambda}$, and the power allocation
matrix ${\bf P}$, the mutual information between ${\bf u}$ and ${\bf y}$ is given by
\begin{eqnarray}
\label{cap4}
&&\hspace{-5mm}I_{\mathcal S}({\bf u};{\bf y}|{\mathbf \Lambda},{\bf P} ,{\bf G}) =
h({\bf y} | {\mathbf \Lambda},{\bf P},{\bf G}) - h({\bf w}) \\ \nonumber
&&\hspace{-3mm}=  -\!\!\int_{{\bf y} \in {\mathbb C}^{n_r}}
\hspace{-4mm} p({\bf y}|{\mathbf \Lambda},{\bf P},{\bf G})\log_2(p({\bf y}|{\mathbf \Lambda},{\bf P},{\bf G})) d{\bf y}  - n\log_2(\pi e )
\end{eqnarray}
where the received vector pdf is given by
\begin{equation}
\label{py2}
p({\bf y}|{\mathbf \Lambda},{\bf P},{\bf G}) = \frac{1}{\vert {\mathcal S} \vert \pi^n}
\sum_{{\bf u} \in {\mathcal S}} e^{-\Vert {\bf y} - \sqrt{P_T}{\bf U}{\mathbf \Lambda} {\bf P} {\bf G} {\bf u} \Vert ^2}
\end{equation}
and when $n = n_r$ (i.e., $n_r \leq n_t$), it is equivalently given by
\begin{equation}
\label{py}
p({\bf y}|{\mathbf \Lambda},{\bf P},{\bf G}) = \frac{1}{\vert {\mathcal S} \vert \pi^n}
\sum_{{\bf u} \in {\mathcal S}} e^{-\Vert {\bf r} - \sqrt{P_T}{\mathbf \Lambda} {\bf P} {\bf G} {\bf u} \Vert ^2}
\end{equation}
where ${\bf r} = (r_1, r_2, \cdots, r_n)^T \Define {\bf U}^\dag{\bf y}$.

We next define the capacity of the MIMO Gaussian channel when precoding with  ${\bf G}$.
In the following, we assume that $n_r \leq n_t$, so that $I_{\mathcal S}({\bf u};{\bf y}|{\mathbf \Lambda},{\bf P} ,{\bf G})  =  I_{\mathcal S}({\bf u};{\bf r}|{\mathbf \Lambda},{\bf P} ,{\bf G})$.
Note that, when $n_r > n_t$, the receiver processing
${\bf r} = {\bf U}^\dag{\bf y}$ becomes information lossy,
and $I_{\mathcal S}({\bf u};{\bf y}|{\mathbf \Lambda},{\bf P} ,{\bf G})
  >  I_{\mathcal S}({\bf u};{\bf r}|{\mathbf \Lambda},{\bf P} ,{\bf G})$.

We introduce the following definitions.
For a given pairing $\ell$, let ${\bf r}_k \Define ( r_{i_k}, r_{j_k} )^T$,
${\bf u}_k \Define ( u_{i_k} , u_{j_k} )^T$,
${\mathbf \Lambda}_k \Define \mbox{diag}(\lambda_{i_k}, \lambda_{j_k})$,
${\bf P}_k \Define \mbox{diag}( p_{i_k}, p_{j_k} )$ and
${\mathcal S}_k \Define {\mathcal U}_{i_k} \times {\mathcal U}_{j_k}$.
Due to the pairing structure of ${\bf G}$ the mutual information
$I_{\mathcal S}({\bf u};{\bf r}|{\mathbf \Lambda},{\bf P} ,{\bf G})$
can be expressed as the sum of mutual information of all the $n/2$ pairs as follows:
\begin{eqnarray}
\label{cap4_prl}
%%I_{\mathcal S}({\bf u};{\bf y}|{\mathbf \Lambda},{\bf P} ,{\bf G}) %%&=&
I_{\mathcal S}({\bf u};{\bf r}|{\mathbf \Lambda},{\bf P} ,{\bf G})
&=& \sum_{k = 1}^{n/2} I_{{\mathcal S}_k}({\bf u}_k;{\bf r}_k |{\mathbf \Lambda}_k,{\bf P}_k ,{\theta_k})
\end{eqnarray}

Having fixed the precoder structure to ${\bf T} =  {\bf V}^\dag {\bf P}{\bf G}$, we can formulate the following
\begin{problem}\label{cap_discrete_Xcoded}
\begin{eqnarray}
\label{cap_Xcoded}
C_X({\bf H},P_T) =  \max_{{\bf G}, {\bf P}  \,| \,
   {\bf u} \in {\mathcal S}, \mbox{tr}({\bf P}{\bf P}^\dag) = 1}   I_{\mathcal S}({\bf u};{\bf r}|{\mathbf \Lambda},{\bf P} ,{\bf G})
\end{eqnarray}
\end{problem}
It is clear that the solution of the above problem is still a formidable task, although it is simpler than Problem \ref{cap_discrete}.
In fact, instead of the $n\times n$ variables of ${\bf T}$,  we now deal with  $n$ variables for power allocation in ${\bf P}$,
$n/2$ variables for the angles defining ${\bf A}_k$, and the pairing $\ell\in {\cal L}$. In the following, we will show how to
efficiently solve Problem \ref{cap_discrete_Xcoded} by splitting it into two simpler problems.
%%which can be easily solved.

Power allocation can be divided into power allocation among the $n/2$ pairs, followed by power allocation between the two subchannels of each pair.
Let ${\bar {\bf P}} = \mbox{diag}({\bar p_1} , {\bar p_2}, \cdots, {\bar p_{n/2}})$ be a diagonal matrix,
where ${\bar p_k} \Define \sqrt{p_{i_k}^2 + p_{j_k}^2}$ with ${\bar p_k}^2$ being the power allocated to the $k$-th pair.
The power allocation within each pair can be simply expressed in terms of the fraction $f_k \Define p_{i_k}^2 / {\bar p_k}^2$
of the power assigned to the first subchannel of the pair. The mutual information achieved by the $k$-th pair is then given by
\begin{eqnarray}
\label{cap4_prl1}
&&\hspace{-15mm}I_{{\mathcal S}_k}({\bf u}_k;{\bf r}_k |{\mathbf \Lambda}_k,{\bf P}_k ,{\theta_k})
   =  I_{{\mathcal S}_k}({\bf u}_k;{\bf r}_k |{\mathbf \Lambda}_k,{\bar p_k},f_k ,{\theta_k}) \\
&&\hspace{-6mm} =  -\int_{ {\bf r}_k \in {\mathbb C}^2 }
                    p({\bf r}_k)\log_2 p({\bf r}_k) \, d{\bf r}_k   - 2\log_2(\pi e )  \nonumber
\end{eqnarray}
where $p({\bf r}_k)$ is given by
\begin{equation}
\label{pdfrk}
p({\bf r}_k)  = \frac{1}{\vert {\mathcal S}_k \vert \pi^2}
\sum_{{\bf u}_k \in {\mathcal S}_k} e^{-\Vert {\bf r}_k -  \sqrt{P_T}{\bar p_k}{\mathbf \Lambda}_k {\bf F}_k {\bf A}_k {\bf u}_k \Vert ^2}
\end{equation}
where ${\bf F}_k \Define \mbox{diag}(\sqrt{f_k}, \sqrt{1 - f_k})$ and ${\bf A}_k$ is given by (\ref{akmat}).

The capacity of the discrete input MIMO Gaussian channel when precoding with X-Codes
%%(Problem \ref{cap_discrete_Xcoded})
%%can  be approximated by solving
can be expressed as
\begin{problem}\label{capxcodesapprox}
\begin{eqnarray}
\label{cap_xcodes_approx}
{C}_{X}({\bf H},P_T) = \max_{\ell \in {\mathcal L}, {\bar {\bf P}} |
\mbox{tr}({\bar {\bf P}}{\bar {\bf P}}^\dag) = 1} \sum_{k = 1}^{n/2} C_{{\mathcal S}_k}(k,\ell,{\bar p_k})
\end{eqnarray}
\end{problem}
where $C_{{\mathcal S}_k}(k,\ell,{\bar p_k})$, the capacity of the $k$-th pair
in the pairing $\ell$, is achieved by solving
\begin{problem}\label{capxcodes_pair}
\begin{eqnarray}
\label{cap_xcodes_pair}
C_{{\mathcal S}_k}(k,\ell,{\bar p_k}) =
\max_{\theta_k,f_k} I_{{\mathcal S}_k}({\bf u}_k;{\bf r}_k |{\mathbf \Lambda}_k,{\bar p_k},f_k ,{\theta_k})
\end{eqnarray}
\end{problem}

In other words, we have split Problem \ref{cap_discrete_Xcoded} into two different simpler problems.
%%parts to get a local maximum
%%solution ${\tilde C}_{X}({\bf H},P_T) \leq C_X({\bf H},P_T)$.
Firstly, given a pairing $\ell$ and power allocation between pairs ${\bar {\bf P}}$,
we can solve Problem \ref{capxcodes_pair} for each $k = 1,2, \cdots, n/2$.
Problem \ref{capxcodesapprox} uses the solution to Problem \ref{capxcodes_pair} to find the optimal
pairing $\ell^{*}$ and the optimal power allocation ${\bar {\bf P}}^{*}$ between the $n/2$ pairs.
%This decomposition results in significant reduction in the complexity of computing the optimal X-Codes.
For small $n$, the optimal pairing and power allocation between pairs can always be
computed numerically and by brute force enumeration of all possible pairings.
This is, however, prohibitively complex for large $n$, and we shall discuss
heuristic approaches in Section \ref{multi_subch}.

We will show in the following that, although suboptimal, precoding with
X-Codes will provide a close to optimal capacity
with the additional benefit that the detection complexity at the receiver
is highly reduced, since there is coupling only
between pairs of channels, as compared to the case of full-coupling for
the optimal precoder in \cite{cruz}.

In the next section, we solve Problem \ref{capxcodes_pair}, which is equivalent to finding the optimal
rotation angle and power allocation for a Gaussian MIMO channel with only $n = 2$ subchannels.

%%%%%%%%%%%%%%%%%%%%%%%%%%%%%%%%%%%%%%%%%%%%%%%%%%%%%%%%%%%%%%%%%%%%%%%%%%%%%%%%%%%%%%%%%%%%%%%%%%
\section{Gaussian MIMO channels with $n = 2$}\label{two_subch}
%%%%%%%%%%%%%%%%%%%%%%%%%%%%%%%%%%%%%%%%%%%%%%%%%%%%%%%%%%%%%%%%%%%%%%%%%%%%%%%%%%%%%%%%%%%%%%%%%%

With $n = 2$, there is only one pair and only one possible pairing.
Therefore, we drop the subscript $k$ in Problem  \ref{capxcodes_pair} and we find $C_X({\bf H},P_T)$ in Problem \ref{cap_discrete_Xcoded}.
The processed received vector ${\bf r} \in {\mathbb C}^2$ is given by
\begin{equation}
\label{rxn2}
{\bf r} = \sqrt{P_T} {\mathbf \Lambda} {\bf F} {\bf A} {\bf u} + {\bf z}
\end{equation}
where ${\bf z} = {\bf U}^\dag{\bf w}$ is the equivalent noise vector with the same statistics as ${\bf w}$.
Let $\alpha \Define \lambda_1^2 + \lambda_2^2$ be the overall channel power gain and
$\beta \Define \lambda_1/\lambda_2$ be the {\em condition number} of the channel.
Then (\ref{rxn2}) can be re-written as
\begin{equation}
\label{rxn3}
{\bf r} = \sqrt{{\Tilde P_T}} {\mathbf {\Tilde \Lambda}} {\bf F} {\bf A} {\bf u} + {\bf z}
\end{equation}
where ${\Tilde P_T} \Define P_T \alpha$ and ${\mathbf {\Tilde \Lambda}} \Define
{\mathbf \Lambda}/ \sqrt{\alpha} = \mbox{diag}( \beta/\sqrt{1 + \beta^2}, 1/\sqrt{1 + \beta^2})$.
The equivalent channel ${\mathbf {\Tilde \Lambda}}$ now has a gain of $1$,
and its channel gains are dependent only upon $\beta$.
Our goal is, therefore, to find the optimal rotation angle $\theta^{*}$
and the fractional power allocation $f^{*}$,
which maximize the mutual information of the equivalent channel with
condition number $\beta$ and gain $\alpha=1$.
The total available transmit power is now ${\Tilde P_T}$.

It is difficult to get analytic expressions for the optimal $\theta^{*}$ and $f^{*}$, and therefore
we can use numerical techniques to evaluate them and store them in lookup tables to be used at run time.
%However, since this is not a practical approach,
%we next discuss a practical scheme to achieve mutual information close to capacity.
For a given application scenario, given the distribution of $\beta$, we decide upon a few discrete
values of $\beta$ which are representative of the actual values observed in real channels.
For each such quantized value of $\beta$,
we numerically compute a table of the optimal values $f^{*}$ and $\theta^{*}$ as a function of ${\Tilde P_T}$.
These tables are constructed offline. During the process of communication, the transmitter knows the
value of $\alpha$ and $\beta$ from channel measurements.
It then finds the lookup table with the closest value of $\beta$ to the measured one.
The optimal values $f^{*}$ and $\theta^{*}$ are then found by indexing the appropriate entry in
the table with ${\Tilde P_T}$ equal to $P_T \alpha$.

%----------------------------------------------------------
\begin{figure}[t]
\begin{center}
\hspace{-1mm}
\epsfig{file=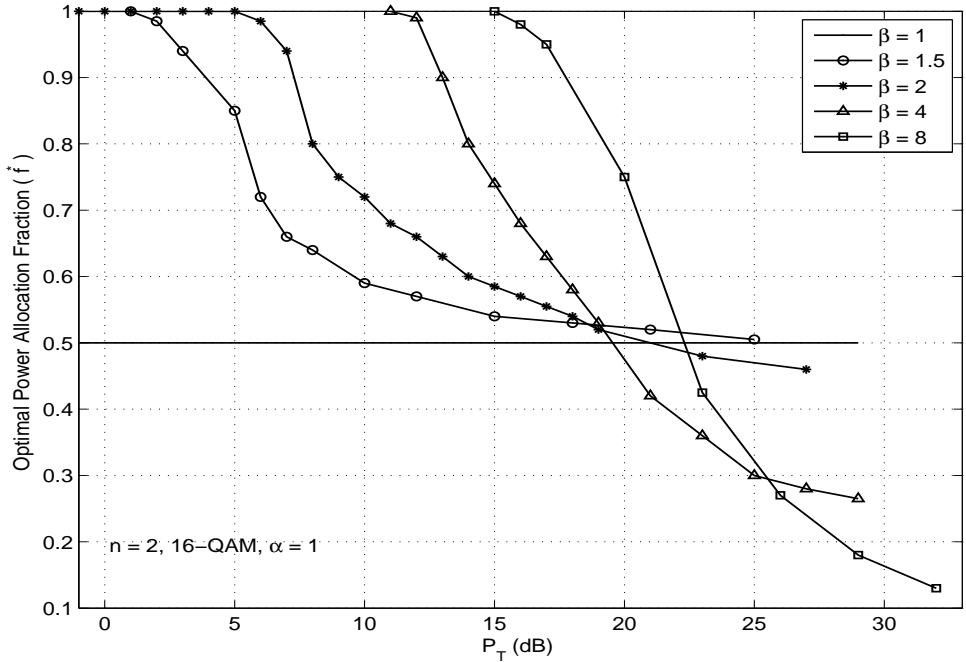, width=130mm,height=90mm}
\end{center}
\vspace{-3mm}
\caption{Plot of $f^*$ versus $P_T$ for $n=2$ parallel channels with $\beta=1,1.5,2,4,8$ and $\alpha=1$. Input alphabet is $16$-QAM.}
\label{palloc_frac}
\end{figure}
%----------------------------------------------------------
In Fig.~\ref{palloc_frac}, we graphically plot the optimal power fraction $f^*$
to be allocated to the stronger channel in the pair, as a function of $P_T$.
The input alphabet is 16-QAM and $\beta = 1,1.5,2,4,8$.
For $\beta = 1$, both channels have equal gains, and therefore, as expected, the optimal power
allocation is to divide power equally between the two subchannels.
However with increasing $\beta$, the power allocation becomes more asymmetrical.
It is observed that at low $P_T$ it is optimal to allocate all power to the stronger channel.
At high $P_T$ the opposite is true, and it is the weaker channel which gets most of the power.
For a fixed $\beta$, as $P_T$ increases, the power allocated to the stronger channel is
shifted to the weaker channel. For a fixed $P_T$, a higher fraction of the
total power is allocated to the weaker channel with increasing $\beta$.
In the high $P_T$ regime, these results are in contrast with the waterfilling scheme,
where almost all subchannels are allocated equal power.

%----------------------------------------------------------
\begin{figure}[t]
\begin{center}
\hspace{-1mm}
\epsfig{file=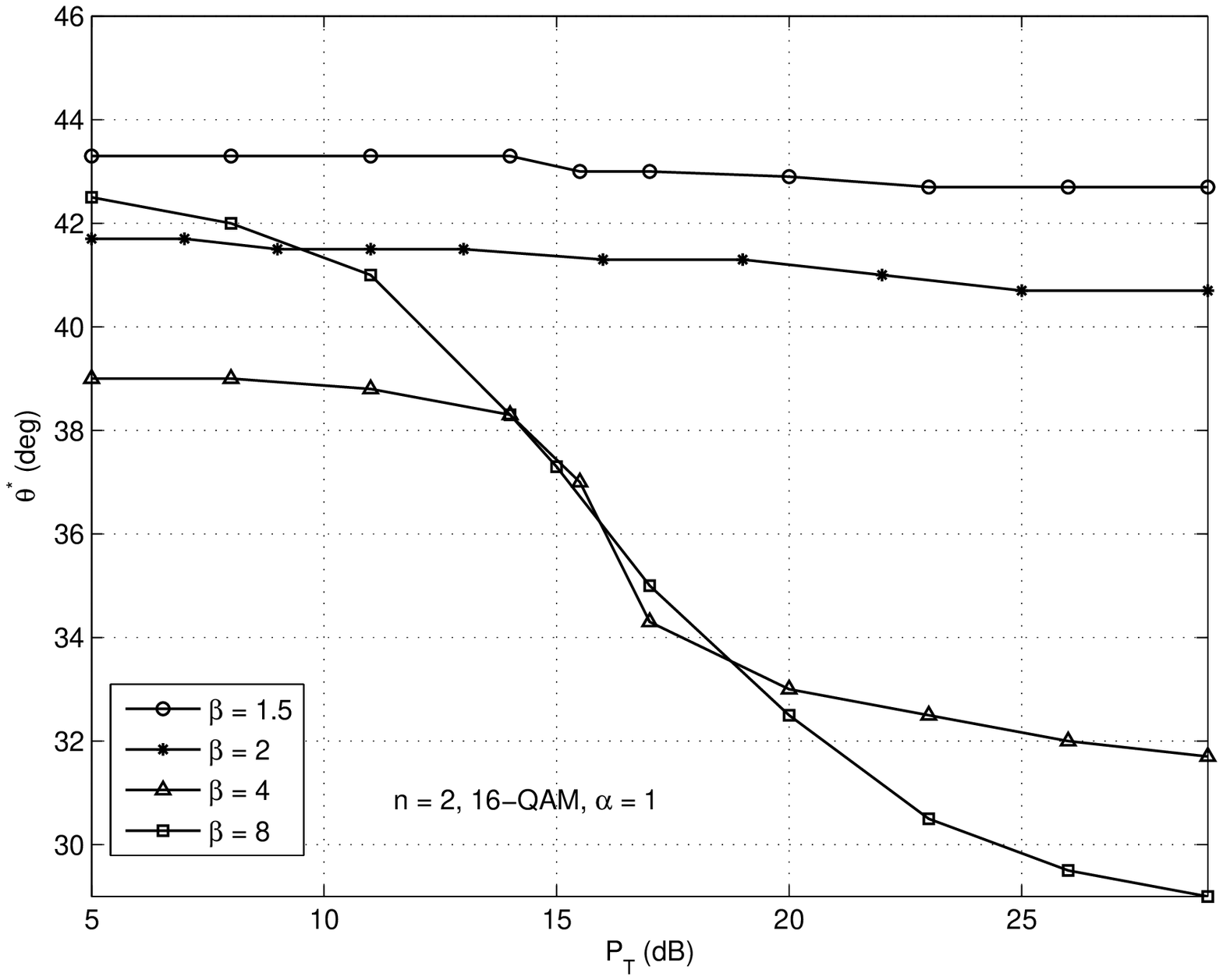, width=130mm,height=90mm}
\end{center}
\vspace{-3mm}
\caption{Plot of $\theta^*$ versus $P_T$ for $n=2$ parallel channels with $\beta=1.5,2,4,8$ and $\alpha=1$. Input alphabet is $16$-QAM. }
\label{palloc_theta}
\end{figure}
%----------------------------------------------------------
In Fig.~\ref{palloc_theta}, the optimal rotation angle $\theta^*$ 
is plotted as a function of $P_T$.
The input alphabet is 16-QAM and $\beta = 1.5,2,4,8$.
For $\beta = 1$ the mutual information is independent of $\theta$
for all values of $P_T$.
For $\beta = 1.5, 2$, the optimal rotation angle is almost invariant to $P_T$.
For larger $\beta$, the optimal rotation angle varies with $P_T$
and approximately ranges between $30-40^\circ$ for all $P_T$ values of interest.

%----------------------------------------------------------
\begin{figure}[t]
\begin{center}
\hspace{-1mm}
\epsfig{file=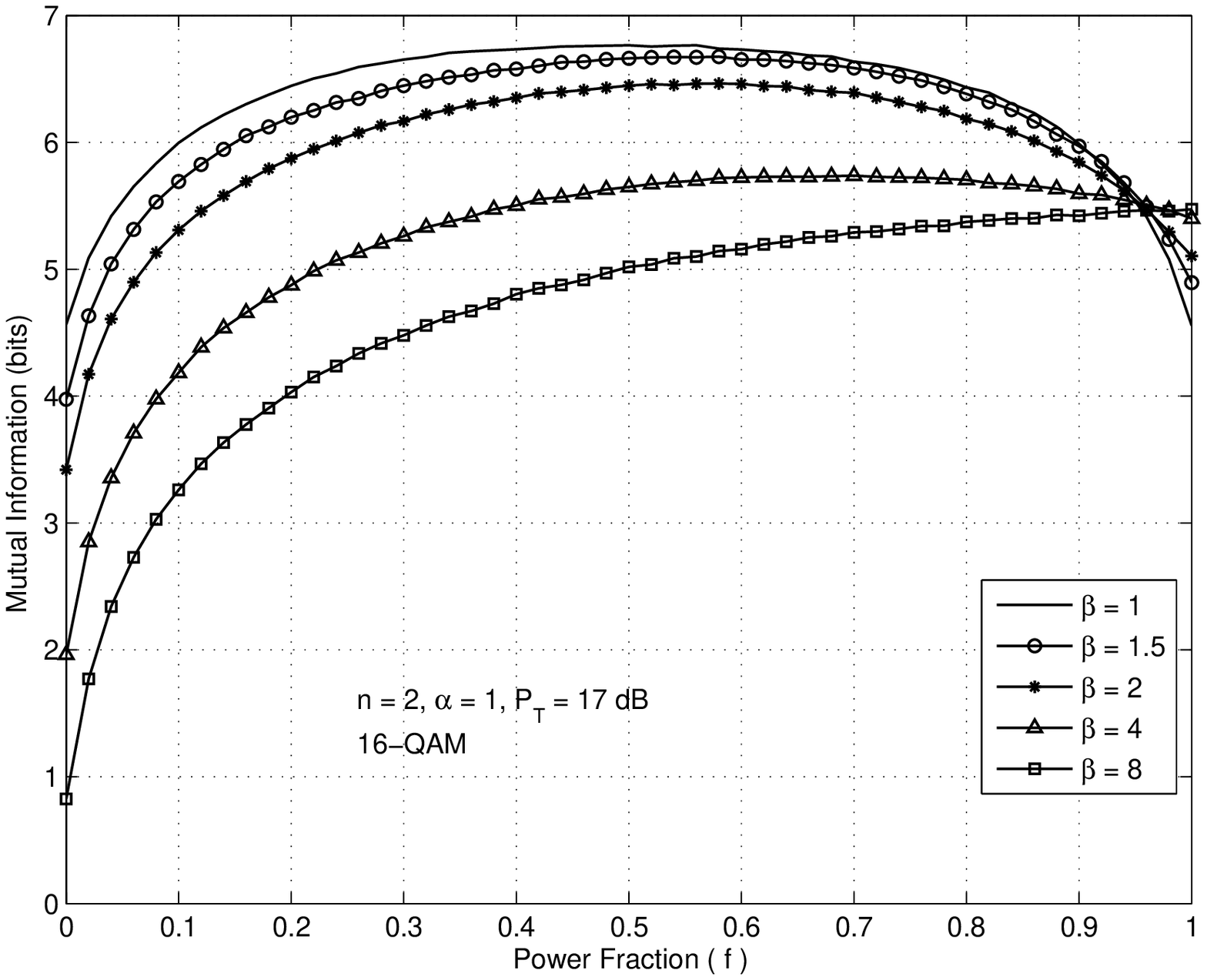, width=130mm,height=90mm}
\end{center}
\vspace{-3mm}
\caption{Mutual Information of X-Codes versus power allocation fraction $f$ for
$n=2$ parallel channels with $\beta=1,1.5,2,4,8$, $\alpha=1$ and $P_T$ = 17 dB. Input alphabet is $16$-QAM.}
\label{MI_vs_pfrac_17dB_16qam}
\end{figure}
%----------------------------------------------------------
Fig.~\ref{MI_vs_pfrac_17dB_16qam} shows the variation of the mutual
information with the power fraction $f$ for $\alpha = 1$.
The power $P_T$ is fixed at 17 dB and the input alphabet is 16-QAM.
We observe that for all values of $\beta$, the mutual information is a concave
function of  $f$. We also observe that
the sensitivity of the mutual information to variation in $f$ increases with increasing $\beta$.
However, for all $\beta$, the mutual information is fairly stable (has a ``plateau'') around the optimal power fraction.
This is good for practical implementation, since this implies that an error in choosing the correct
power allocation would result in a very small loss in the achieved mutual information.

%----------------------------------------------------------
\begin{figure}[t]
\begin{center}
\hspace{-1mm}
\epsfig{file=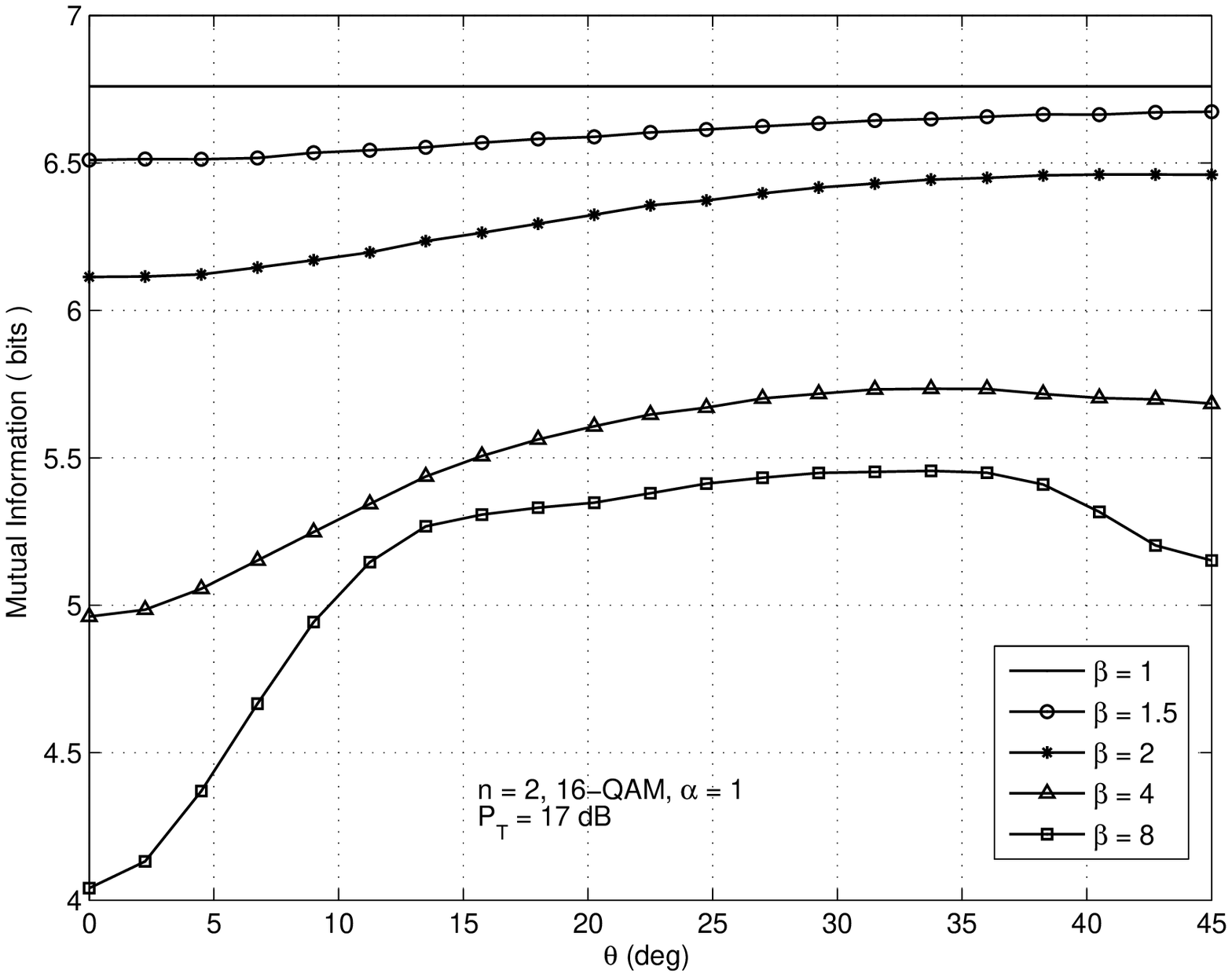, width=130mm,height=90mm}
\end{center}
\vspace{-3mm}
\caption{Mutual information of X-Codes versus rotation angle $\theta$ for $n=2$
parallel channels with $\beta=1,1.5,2,4,8$, $\alpha=1$ and $P_T$ = 17 dB. Input alphabet is $16$-QAM.}
\label{MI_vs_theta_17dB_16qam}
\end{figure}
%----------------------------------------------------------
In Fig.~\ref{MI_vs_theta_17dB_16qam}, we plot the variation of the mutual information w.r.t.
the rotation angle $\theta$. The power $P_T$ is fixed at 17 dB and the input alphabet is 16-QAM. For $\beta = 1$, the mutual information is obviously constant with $\theta$.
With increasing $\beta$, mutual information is observed to be increasingly sensitive to $\theta$.
However, when compared with Fig.~\ref{MI_vs_pfrac_17dB_16qam}, it can also be seen that the mutual
information appears to be more sensitive to the power allocation fraction $f$, than to $\theta$.

%----------------------------------------------------------
\begin{figure}[t]
\begin{center}
\hspace{-1mm}
\epsfig{file=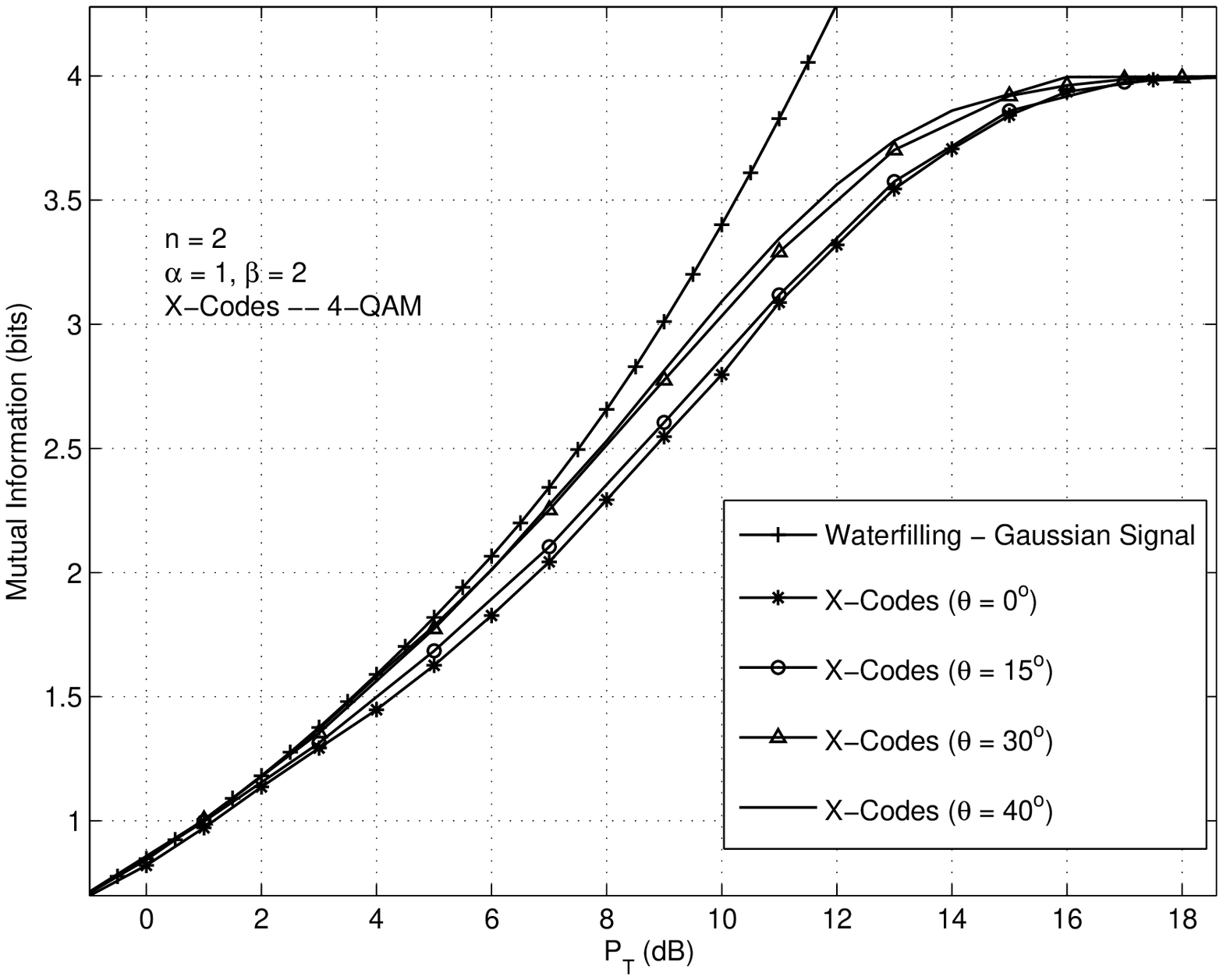, width=130mm,height=90mm}
\end{center}
\vspace{-3mm}
\caption{Mutual information versus $P_T$ for X-Codes for different $\theta$s,
$n=2$ parallel channels, $\alpha=1$, $\beta=2$, and 4-QAM input alphabet.}
\label{thetavar_4qam}
\end{figure}
%----------------------------------------------------------
In Fig.~\ref{thetavar_4qam}, we plot the mutual information of X-Codes
for different rotation angles with $\alpha = 1$ and $\beta = 2$.
For each rotation angle, the power allocation is optimized numerically.
%For a given $\alpha$ and $\beta$,
We observe that, the mutual information is quite sensitive to the
rotation angle except in the range 30-40$^\circ$.
%We also observed that the optimal angle ${\theta}^{*}$ does not vary %much with $P_T$.

We next present some simulation results to show that indeed our simple precoding scheme
can significantly increase the mutual information, compared to the case of no precoding across
subchannels (i.e., Mercury/waterfilling).
For the sake of comparison, we also present the mutual information achieved by the waterfilling scheme
with discrete input alphabets.
%Recall that waterfilling is only optimal when inputs are Gaussian distributed,
%but this is, however, not true when input alphabet is discrete.

We restrict the discrete input alphabets ${\mathcal U}_i, i=1,2$,
to be square $M$-QAM alphabets consisting of two $\sqrt{M}$-PAM alphabets in quadrature.
Mutual information is evaluated by solving Problem \ref{capxcodes_pair}
(i.e., numerically maximizing w.r.t. the rotation angle and power allocation).

%----------------------------------------------------------
\begin{figure}[t]
\begin{center}
\hspace{-1mm}
\epsfig{file=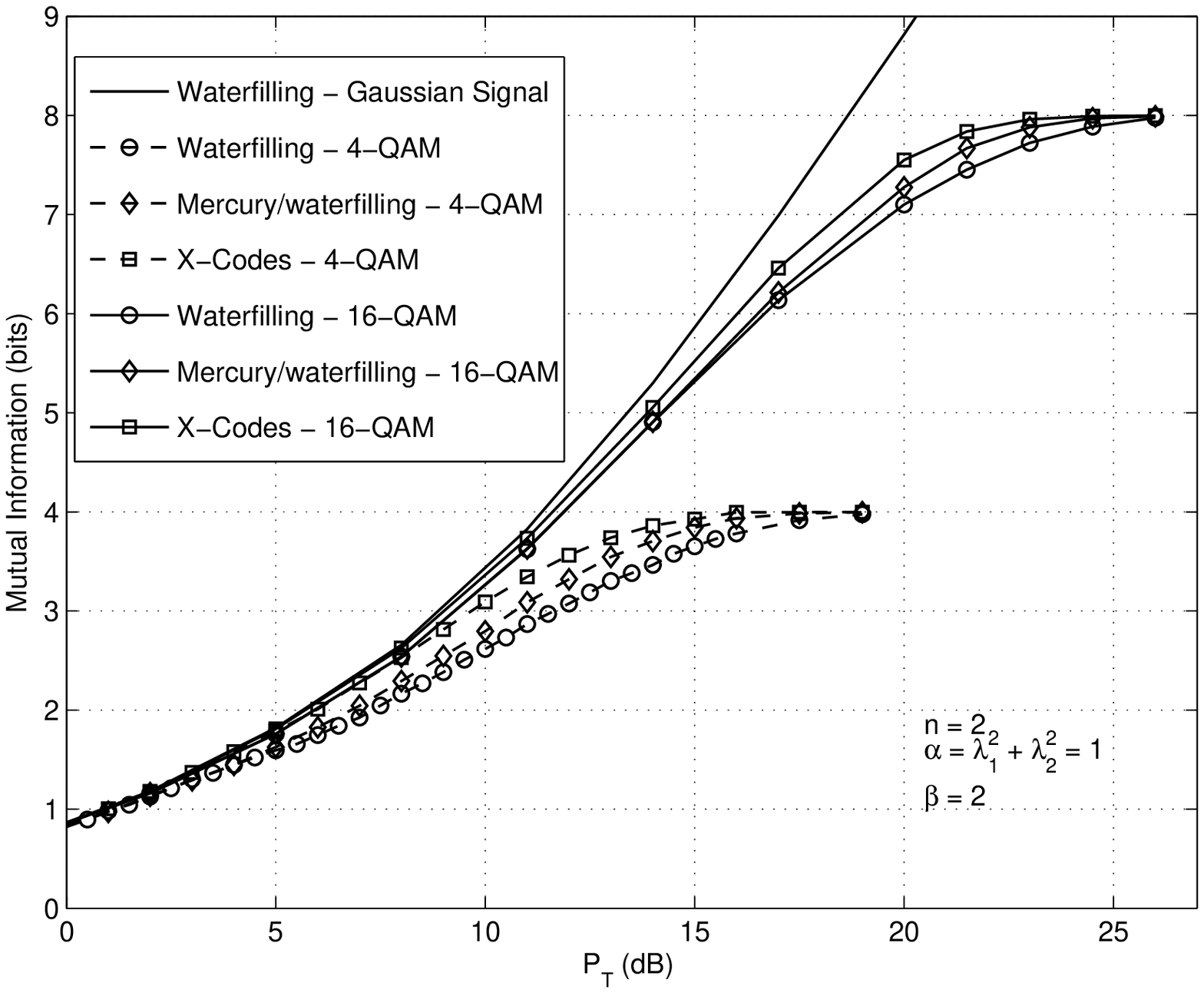, width=130mm,height=90mm}
\end{center}
\vspace{-3mm}
\caption{Mutual information versus $P_T$ for $n=2$ parallel channels with $\beta=2$ and $\alpha=1$, for 4-QAM and 16-QAM.}
\label{beta2_4_16qam}
\end{figure}
%----------------------------------------------------------
In Fig.~\ref{beta2_4_16qam}, we plot the maximal mutual information versus $P_T$, for a system with two subchannels, $\beta=2$ and $\alpha = 1$.
Mutual information is plotted for 4- and 16-QAM signal sets.
It is observed that  for a given achievable mutual information,
coding across subchannels is more power efficient.
For example, with 4-QAM and an achievable mutual information of $3$ bits, X-Codes require only $0.8$ dB more transmit power when
compared to the ideal Gaussian signalling with waterfilling. This gap increases to $1.9$ dB for Mercury/waterfilling
and $2.8$ dB for the waterfilling scheme with $4$-QAM as the input alphabet. A similar trend is observed with $16$-QAM as the input alphabet.
The proposed precoder clearly performs better, since the mutual information is optimized
w.r.t. the rotation angle $\theta$ and power allocation, while Mercury/waterfilling,
as a special case of X-Code, only optimizes power allocation and fixes $\theta=0$.

%----------------------------------------------------------
\begin{figure}[t]
\begin{center}
\hspace{-1mm}
\epsfig{file=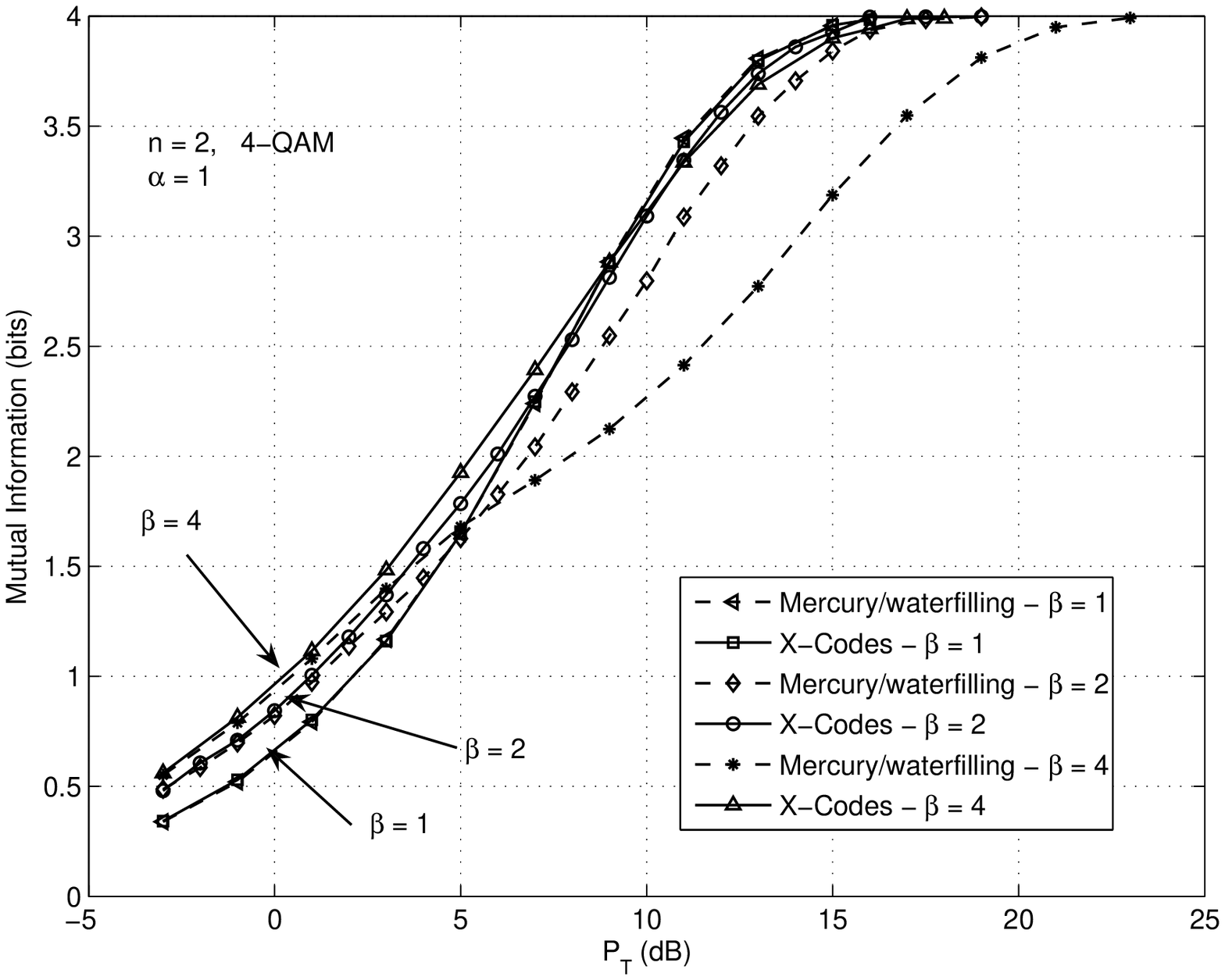, width=130mm,height=90mm}
\end{center}
\vspace{-3mm}
\caption{Mutual information versus $P_T$ for $n=2$ parallel channels with varying $\beta=1,2,4$, $\alpha=1$ and 4-QAM input alphabet.}
\label{beta124_4qam}
\end{figure}
%----------------------------------------------------------
In Fig.~\ref{beta124_4qam}, we compare the mutual information achieved by X-Codes and the
Mercury/waterfilling strategy for $\alpha = 1$ and $\beta=1,2,4$. The input alphabet is $4$-QAM.
It is observed that both the schemes have the same mutual information when $\beta = 1$.
However with increasing $\beta$, the mutual information of Mercury/waterfilling strategy is
observed to degrade significantly at high $P_T$, whereas the performance of X-Codes
does not vary as much.
%For example, for a target mutual information of $3$ bits, with $\beta = 4$ the Mercury/waterfilling
%strategy requires $4.6$ dB more transmit power as compared to when $\beta = 1$.
%On the other hand, X-Codes require only $0.1$ dB extra transmit power
%when $\beta$ increases from $1$ to $4$.
The degradation of mutual information for the Mercury/waterfilling strategy is explained as follows.
For the Mercury/waterfilling strategy, with increasing $\beta$, all the available
power is allocated to the stronger channel till a certain transmit power threshold.
However, since finite signal sets are used, mutual information is bounded from above until
the transmit power exceeds this threshold.
This also explains the reason for the intermediate change of slope in the mutual information
curve with $\beta=4$ (see the rightmost curve in Fig.~\ref{beta124_4qam}).
On the other hand, due to coding across subchannels, this problem does not arise when precoding with X-Codes.
Therefore, in terms of achievable mutual information, rotation coding is observed
to be more robust to ill-conditioned channels.

For low values of $P_T$, mutual information of both the schemes are similar,
and improves with increasing $\beta$.
This is due to the fact that, at low $P_T$, mutual information increases linearly with $P_T$, and
therefore all power is assigned to the stronger channel. With increasing $\beta$, the stronger
channel has an increasing fraction of the total channel gain, which results in increased mutual information.

%----------------------------------------------------------
\begin{figure}[t]
\begin{center}
\hspace{-1mm}
\epsfig{file=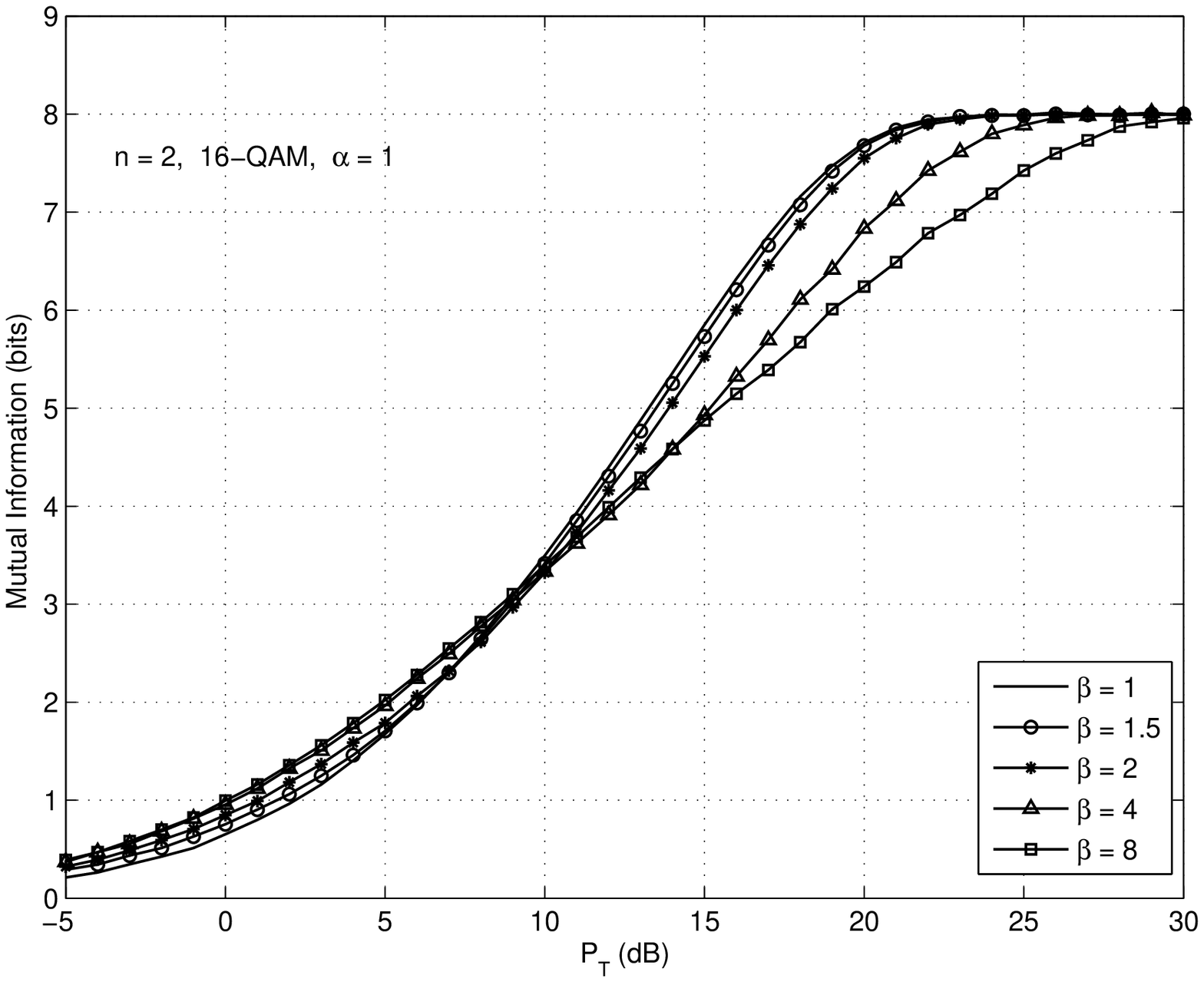, width=130mm,height=90mm}
\end{center}
\vspace{-3mm}
\caption{Mutual information with X-Codes versus $P_T$ for $n=2$ parallel channels with varying $\beta=1,2,4,8$, $\alpha=1$ and  16-QAM input alphabet.}
\label{beta1248_16qam}
\end{figure}
%----------------------------------------------------------
In Fig.~\ref{beta1248_16qam}, the mutual information with X-Codes is plotted for $\beta = 1,2,4,8$ and with 16-QAM as the input alphabet.
It is observed that at low values of $P_T$, a higher value of $\beta$ is favorable.
However at high $P_T$, with 16-QAM input alphabets, the performance degrades with increasing $\beta$.
This degradation is more significant compared to the degradation observed with 4-QAM input alphabets. Therefore it can be concluded
that the mutual information is more sensitive to $\beta$ with 16-QAM input alphabets as compared to 4-QAM.

%%%%%%%%%%%%%%%%%%%%%%%%%%%%%%%%%%%%%%%%%%%%%%%%%%%%%%%%%%%%%%%%%%%%%%%%%%%%%%%%%%%%%%%%%%%%%%%%%%
\section{Gaussian MIMO channels with $n > 2$}\label{multi_subch}
%%%%%%%%%%%%%%%%%%%%%%%%%%%%%%%%%%%%%%%%%%%%%%%%%%%%%%%%%%%%%%%%%%%%%%%%%%%%%%%%%%%%%%%%%%%%%%%%%%
%----------------------------------------------------------
\begin{figure}[t]
\begin{center}
\hspace{-1mm}
\epsfig{file=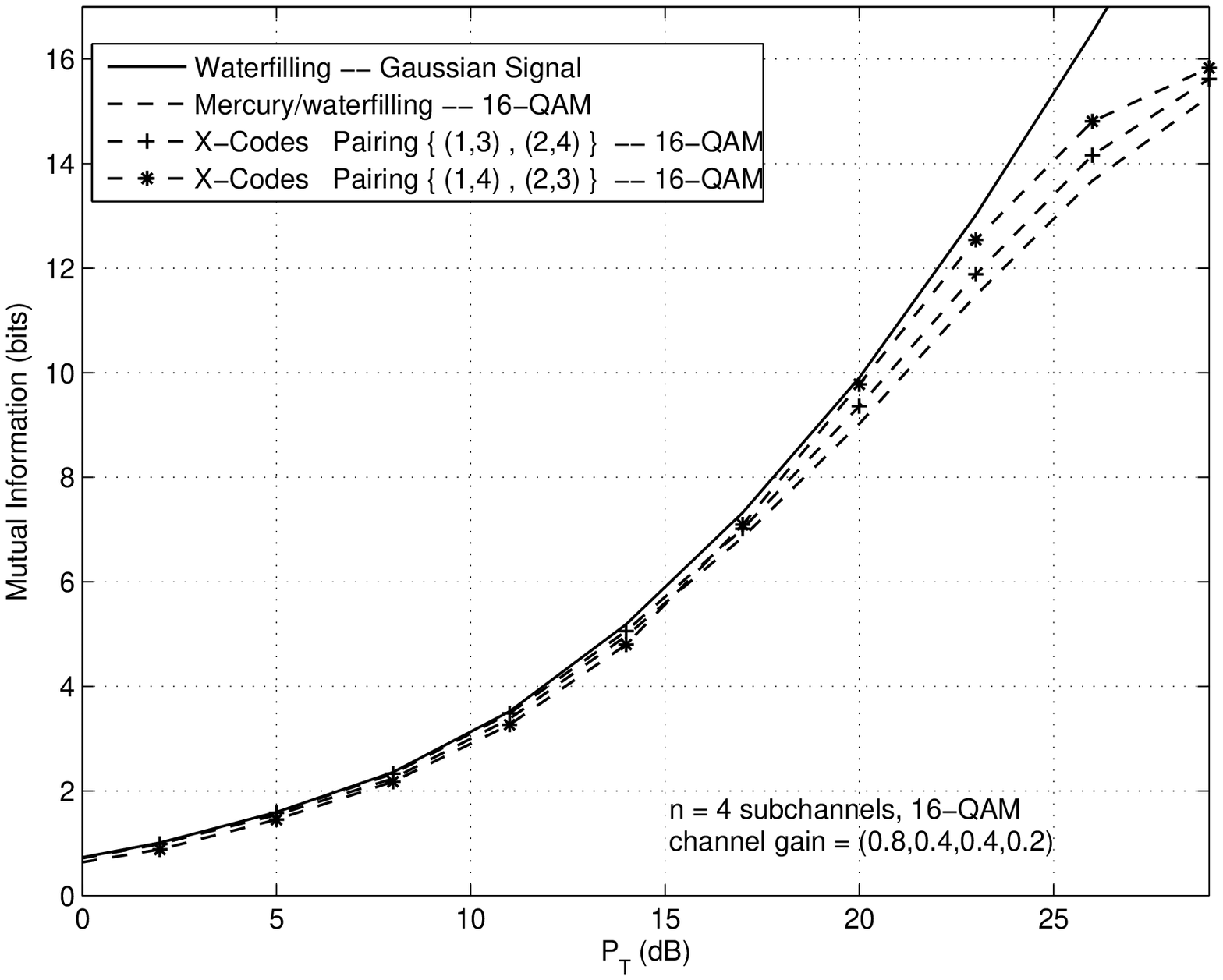, width=130mm,height=90mm}
\end{center}
\vspace{-3mm}
\caption{Mutual information versus $P_T$ with two different pairings for a $n = 4$ diagonal channel and 16-QAM input alphabet.}
\label{n4pair}
\end{figure}
%----------------------------------------------------------
We now consider the problem of finding the optimal pairing and power allocation between pairs for different
Gaussian MIMO channels with even $n$ and $n > 2$.
%We also compare the performance achieved by X-Codes with that of the optimal precoder in \cite{cruz}.
We first observe that mutual information is indeed sensitive to the chosen pairing, and this therefore
justifies the criticality of computing the optimal pairing.
This is illustrated through Fig.~\ref{n4pair}, for $n = 4$ with a diagonal channel ${\mathbf \Lambda} = \mbox{diag}(0.8, 0.4,0.4,0.2)$ and 16-QAM.
Optimal power allocation between the two pairs is computed numerically.
It is observed that the pairing $\{ (1,4), (2,3)\}$ performs significantly better than the pairing $\{ (1,3), (2,4)\}$.

%----------------------------------------------------------
\begin{figure}[t]
\begin{center}
\hspace{-1mm}
\epsfig{file=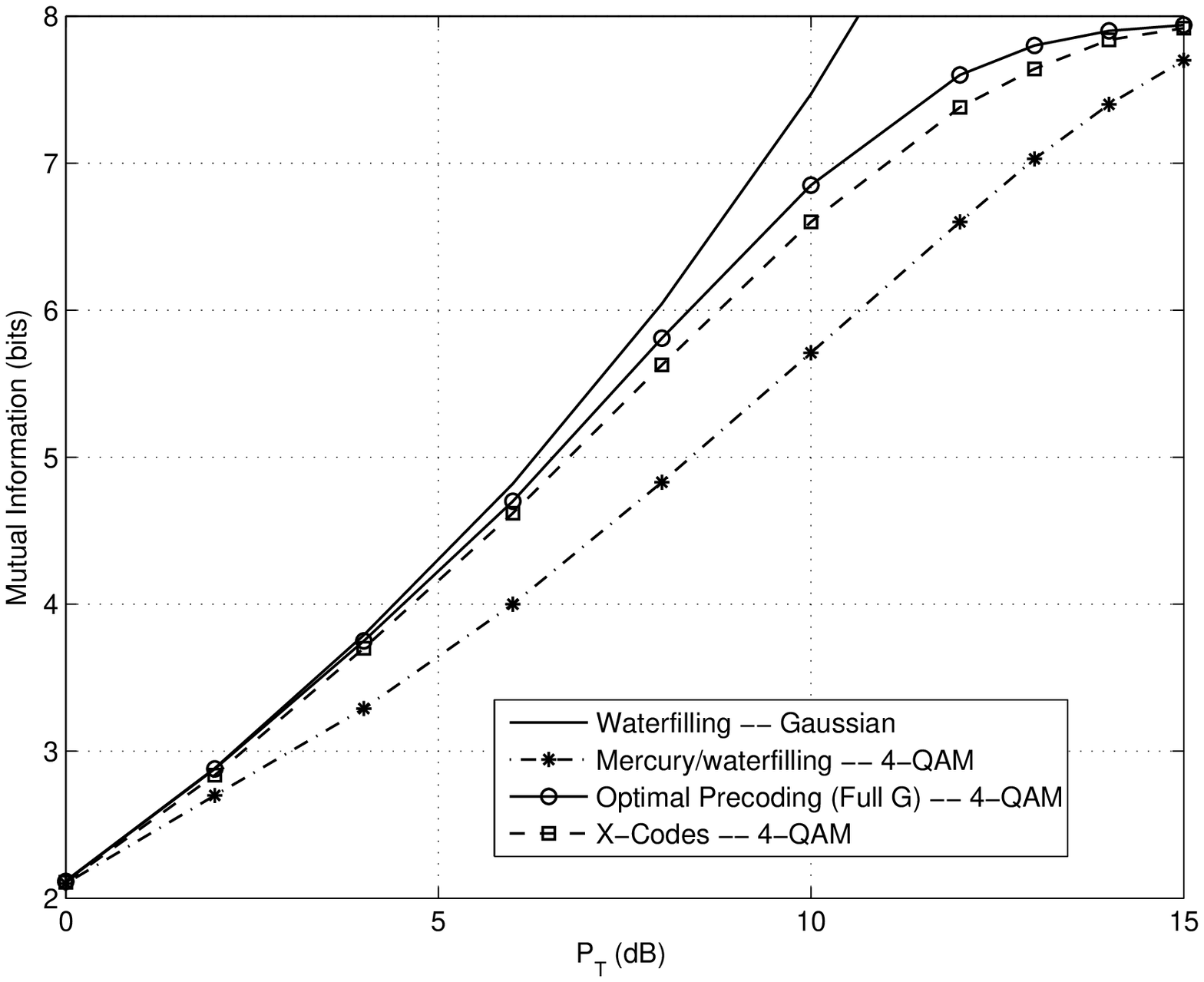, width=130mm,height=90mm}
\end{center}
\vspace{-3mm}
\caption{Mutual information versus $P_T$ for the Gigabit DSL channel given by (42) in \cite{cruz}.}
\label{cruz_cmp}
\end{figure}
%----------------------------------------------------------
In Fig.~\ref{cruz_cmp}, we compare the mutual information achieved with optimal precoding \cite{cruz},
to that achieved by the proposed precoder with 4-QAM input alphabet.
The $4 \times 4$ full channel matrix (non-diagonal channel) is given by (42) in \cite{cruz}.
For X-Codes, the optimal pairing is $\{ (1,4), (2,3)\}$ and the optimal power allocation between the pairs is computed numerically.
It is observed that X-Codes perform very close to the optimal precoding scheme.
Specifically, for an achievable mutual information of 6 bits, compared to the optimal
precoder \cite{cruz}, X-Codes need only 0.4dB extra power whereas 2.3dB extra power is required with Mercury/waterfilling.

%----------------------------------------------------------
\begin{figure}[t]
\begin{center}
\hspace{-1mm}
\epsfig{file=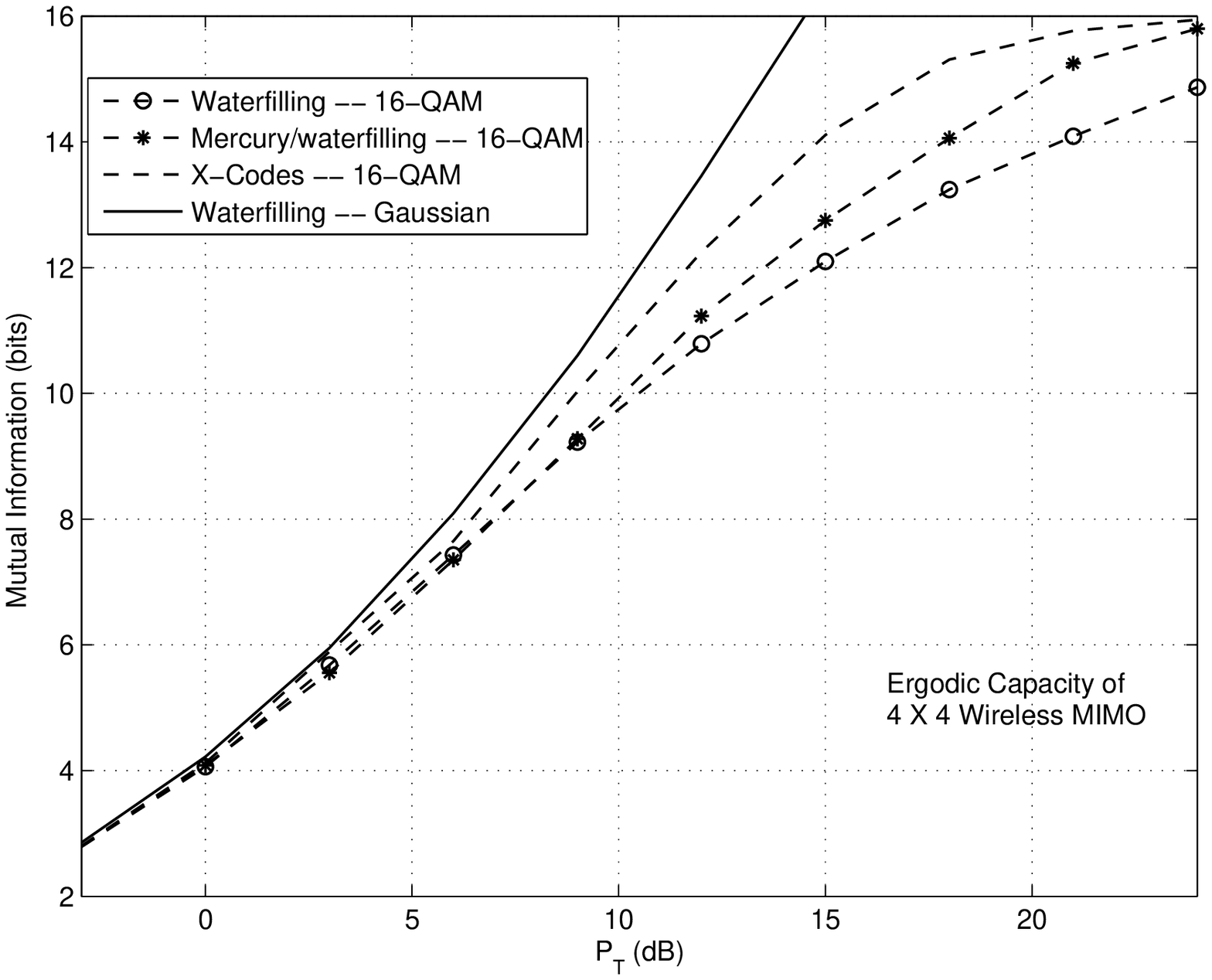, width=130mm,height=90mm}
\end{center}
\vspace{-3mm}
\caption{4 X 4 Wireless MIMO Ergodic capacity.}
\label{mimo_ergodic_cap}
\end{figure}
%----------------------------------------------------------
Another application is in wireless MIMO channels with perfect channel state
information at both the transmitter and receiver.
The channel coefficients are modeled as i.i.d complex normal random variables with unit variance.

In Fig.~\ref{mimo_ergodic_cap}, we plot the ergodic capacity
(i.e., the mutual information averaged over channel realizations)
for a $4 \times 4$ wireless MIMO channel.
For X-Codes, the best pairing and power allocation between pairs are chosen numerically
using the optimal $\theta$ and power fraction tables created offline.
It is observed that at high $P_T$, simple rotation based coding using X-Codes improves
the mutual information significantly, when compared to Mercury/waterfilling.
%It is also observed that X-Codes are more power efficient than the other schemes considered.
For example, for a target mutual information of 12 bits, X-Codes
perform 1.2dB away from the idealistic Gaussian signalling scheme.
This gap from the Gaussian signalling scheme increases to
3.1dB for the Mercury/waterfilling scheme and to 4.4dB for the
waterfilling scheme with 16-QAM alphabets.

In this application scenario the low complexity of our precoding scheme becomes an essential feature, since
the precoder can be computed on the fly using the look-up tables for each channel realization.

%%%%%%%%%%%%%%%%%%%%%%%%%%%%%%%%%%%%%%%%%%%%%%%%%%%%%%%%%%%%%%%%%%%%%%%%%%%%%%%%%%%%%%%%%%%%%%%%%%
\section{Application to OFDM} \label{sec_ofdm}
%%%%%%%%%%%%%%%%%%%%%%%%%%%%%%%%%%%%%%%%%%%%%%%%%%%%%%%%%%%%%%%%%%%%%%%%%%%%%%%%%%%%%%%%%%%%%%%%%%

In OFDM applications, $n$ is large and Problem \ref{capxcodesapprox} becomes too complex to solve,
since we can no more find the optimal pairing by enumeration.

It was observed in Section \ref{two_subch}, that for $n = 2$, a larger value of the condition
number $\beta$ leads to a higher mutual information at low values of $P_T$ (low SNR).
Therefore, we conjecture that pairing the $k$-th subchannel with the $(n/2 + k)$-th subchannel
could have mutual information very close to optimal, since this pairing scheme attempts to maximize the minimum $\beta$ among all pairs.
We shall call this scheme the ``conjectured'' pairing scheme, and the X-Code scheme, which pairs the $k$-th with the $(n-k+1)$-th
subchannel, the ``X-pairing" scheme.
Note that the ``X-pairing'' scheme was proposed in \cite{Xcodes_paper} as a scheme which achieved the optimal diversity gain when precoding with X-Codes.

Given a pairing of subchannels, it is also difficult to compute the optimal power allocation between pairs ${\bar {\bf P}}$.
However, it was observed that for channels with large $n$, even waterfilling power allocation between the
pairs (with ${\alpha}_k \Define \sqrt{ {\lambda}_{i_k}^2 +   {\lambda}_{j_k}^2}$ as the channel gain of the $k$-th pair) results in good performance.

Apart from the ``conjectured'' and the ``X-pairing''
schemes, we propose the following scheme which is based on the ``Hungarian"
assignment algorithm \cite{Kuhn55} and which attempts to find a
good approximation to the optimal pairing.
We shall call this as the ``Hungarian" pairing scheme.
Before describing the ``Hungarian'' pairing scheme, we briefly review the Hungarian assignment problem as follows.

Consider $m$ different workers and $m$ different jobs that have to be completed.
Also let $C(i,j)$ be the cost involved when
the $i$-th worker is assigned to the $j$-th job.
We can therefore think of a cost matrix, whose $(i,j)$-th entry has the value $C(i,j)$.
The Hungarian assignment problem, is to then find the optimal assignment of workers to jobs
(each worker getting assigned to exactly one job) such that the total cost
of getting all the jobs completed is minimized.
It is easy to see, that a maximization job assignment problem could be posed
into a minimization problem and vice versa.

To find a good approximation to the optimal pairing, we split the $n$ subchannels into two groups
{\em i)} Group-I : subchannels 1 to $n/2$, with the $j$-th subchannel in the role of the $j$-th job ($j =1,2, \cdots n/2$),
{\em ii)}  Group-II : subchannels $n/2+1$ to $n$, with the $(n/2+i)$-th subchannel
in the role of the $i$-th worker ($i=1,2, \cdots n/2$). Therefore, there are $n/2$ workers and jobs.

For a given SNR $P_T$, we initially assume uniform power allocation between
all pairs and therefore assign a power of $2P_T/n$ to each pair.
The value of $C(i,j)$ is evaluated by finding the optimal mutual information achieved by an
equivalent $n=2$ channel with the $n/2+i$-th and the $j$-th subchannels as its two subchannels.
This can be obtained by first choosing a table (see Section \ref{two_subch}) with the closest value of
$\beta$ to the given $\lambda_j/\lambda_{n/2+i}$, and then indexing the appropriate entry into the table with SNR=$2P_T(\lambda_j^2 + \lambda_{n/2+i}^2)/n$.
The Hungarian algorithm then finds the pairing with the highest mutual information.
%We refer to this pairing as the ``Hungarian" pairing.
Power allocation between the pairs is then achieved through the waterfilling scheme.

It was observed through monte-carlo simulations that, even uniform power allocation between the
subchannels results in almost same mutual information as achieved through waterfilling between pairs.
This can be explained from the fact that by separating into a group of stronger (Group-I) and a group of
weaker channels (Group-II), any pairing would result in all pairs having almost the same channel gain ${\alpha}_k$.
This therefore implies that the optimal power allocation scheme would allocate nearly equal power
to all pairs, which both the uniform and the waterfilling schemes would also do.
Henceforth, it can be conjectured that with the proposed separation of subchannels into 2 groups,
both the uniform and the waterfilling power allocation schemes would have close to optimal performance,
and any further improvement in mutual information by optimizing the power allocation would be minimal.
This also supports the initial usage of uniform power $2P_T/n$ to compute the entries $C(i,j)$ before executing the Hungarian algorithm.
Furthermore, the computational complexity of the Hungarian algorithm is $O(n^3)$ and is therefore practically feasible.

To study the sensitivity of the mutual information to the pairing of subchannels, we also consider a ``Random'' pairing scheme.
In the ``Random" pairing scheme, we first choose a large number ($\approx$ 50) of random pairings.
For each chosen random pairing we evaluate the mutual information (through monte-carlo simulations)
with waterfilling power allocation between pairs.
Finally the average mutual information is computed. This gives us insight into the mean value of
the mutual information w.r.t. pairing.
It would also help us in quantifying the effectiveness of the heuristic pairing schemes discussed above.

%----------------------------------------------------------
\begin{figure}[t]
\begin{center}
\hspace{-1mm}
\epsfig{file=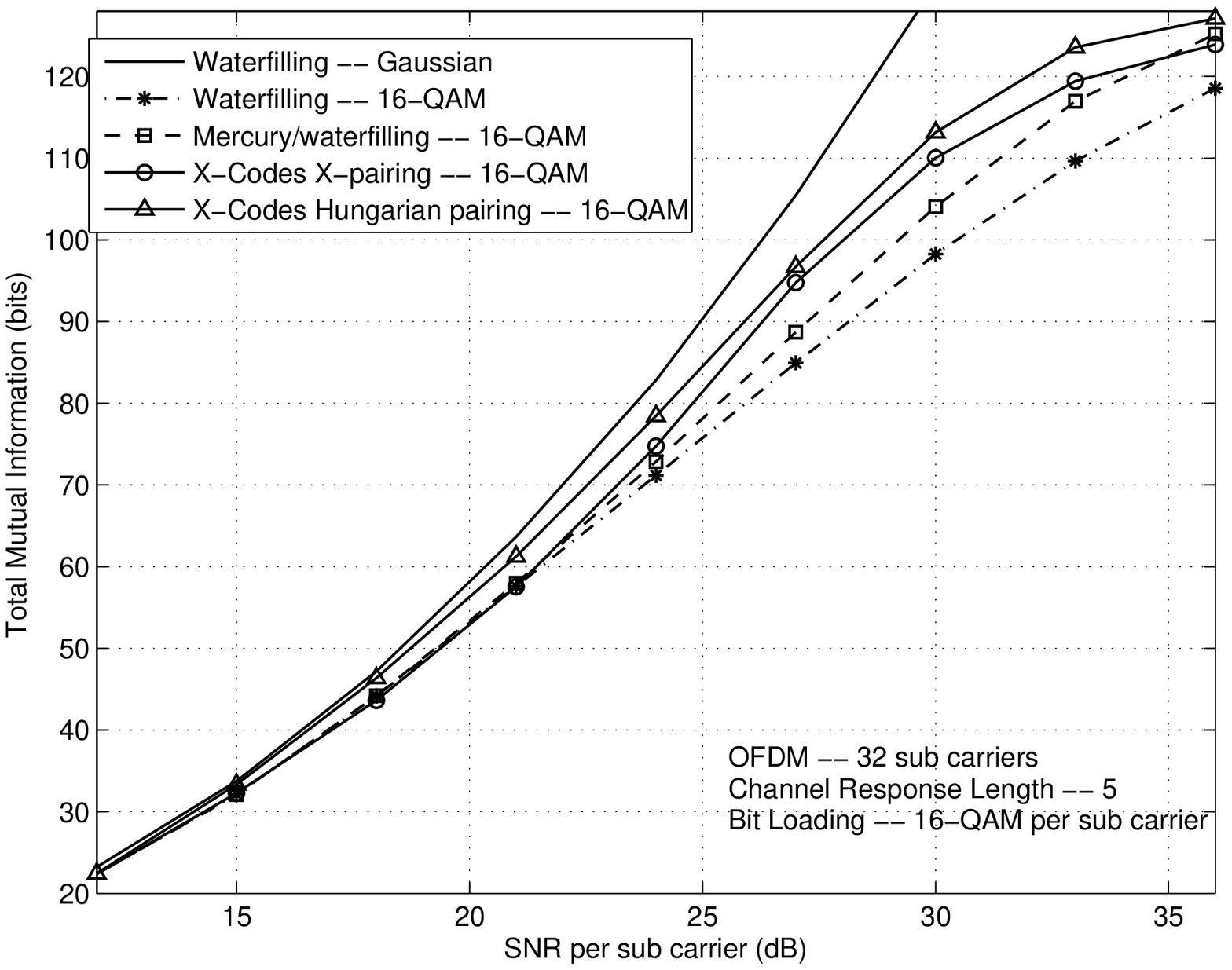, width=130mm,height=90mm}
\end{center}
\vspace{-3mm}
\caption{Mutual information versus per subcarrier SNR for an OFDM system with 32 carriers.
X-Codes versus Mercury/waterfilling. }
\label{ofdm_ex_1}
\end{figure}
%----------------------------------------------------------
%----------------------------------------------------------
\begin{figure}[t]
\begin{center}
\hspace{-1mm}
\epsfig{file=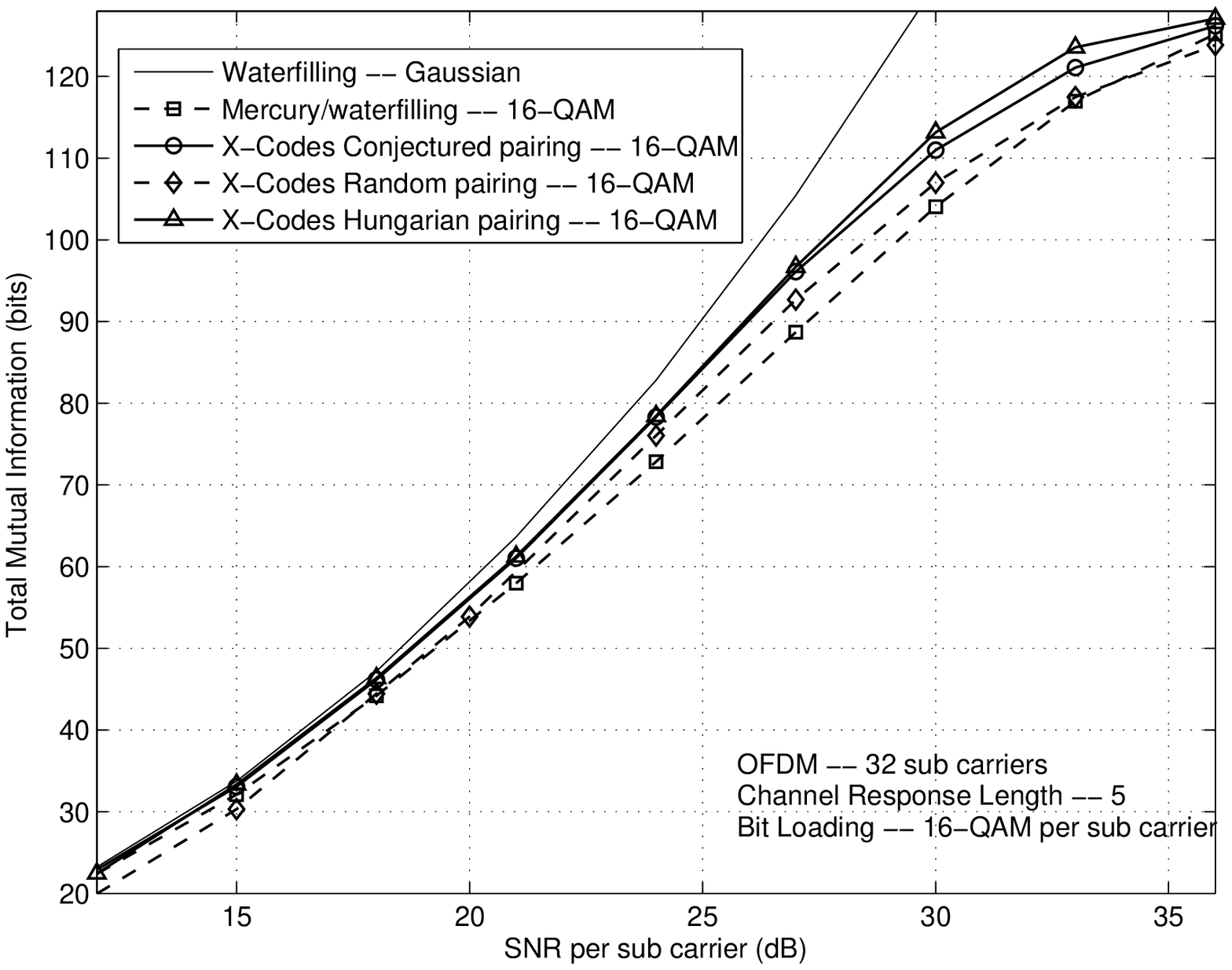, width=130mm,height=90mm}
\end{center}
\vspace{-3mm}
\caption{Mutual information versus per subcarrier SNR for an OFDM system with 32 carriers. Comparison of heuristic pairing schemes.
}
\label{ofdm_ex_2}
\end{figure}
%----------------------------------------------------------

We next illustrate the mutual information achieved by these heuristic schemes for an OFDM system with $n=32$ subchannels and 16-QAM.
The channel impulse response is $[ -0.454 + {\mathfrak j}0.145,  -0.258 + {\mathfrak j}0.198, 0.0783 +
{\mathfrak j}0.069, -0.408 - {\mathfrak j}0.396, -0.532 - {\mathfrak j}0.224 ]$.
For the ``conjectured" and the ``X-pairing" schemes also, power allocation is achieved through waterfilling between the pairs.

In Fig.~\ref{ofdm_ex_1} the total mutual information is plotted
as a function of the SNR per sub carrier. It is observed that the
proposed precoding scheme performs much better than the Mercury/waterfilling scheme.
The proposed precoder with the ``Hungarian" pairing scheme performs within 1.1dB of
the Gaussian signalling scheme for an achievable total mutual information of $96$ bits (i.e., a rate of 96/128 = 3/4).
The proposed precoder with the ``Hungarian" pairing scheme performs about 1.6dB better than the Mercury/waterfilling scheme.
The ``X-pairing" scheme performs better than the Mercury/waterfilling and worse than the ``Hungarian" pairing scheme.
Even at a low rate of 1/2 (i.e., a total mutual information of 64 bits), the proposed precoder with the ``Hungarian" pairing scheme
performs about 0.7dB better than the Mercury/waterfilling scheme.

In Fig.~\ref{ofdm_ex_2}, we compare the mutual information achieved by the various heuristic pairing schemes.
It is observed that the ``conjectured" pairing scheme performs very close
to the ``Hungarian" pairing scheme except at very high SNR.
For example, even for a high mutual information of 96 bits, the ``Hungarian" pairing
scheme performs better than the ``conjectured" pairing scheme by only about 0.2dB.
However at very high rates (like 7/8 and above), the ``Hungarian" pairing scheme is observed to perform better
than the ``conjectured" pairing scheme by about 0.7dB.
Therefore for low to medium rates, it would be better to use the ``conjectured" pairing
since it has the same performance at a lower computational complexity.
The mutual information achieved by the ``Random" pairing scheme is observed to be strictly
inferior than the ``conjectured" pairing scheme at all values of SNR, and at low SNR it is even
worse than the Mercury/waterfilling strategy.
This, therefore implies that the total mutual information is indeed
sensitive to the chosen pairing.
Further, till a rate of 1/2 (i.e., a mutual information of 64 bits) it appears that any extra optimization effort
would not result in significant performance improvement for the ``conjectured"
pairing scheme, since it is already very close to the idealistic Gaussian signalling schemes.
However at higher rate and SNR it may still be possible to improve the mutual information
by further optimizing the selection of pairing scheme and power allocation between pairs.
This is however a difficult problem that requires further investigation.

%%%%%%%%%%%%%%%%%%%%%%%%%%%%%%%%%%%%%%%%%%%%%%%%%%%%%%%%%%%%%%%%%%%%%%%%%%%%%%%%%%%%%%%%%%%%%%%%%%
\section{Conclusions}\label{conclusions}
%%%%%%%%%%%%%%%%%%%%%%%%%%%%%%%%%%%%%%%%%%%%%%%%%%%%%%%%%%%%%%%%%%%%%%%%%%%%%%%%%%%%%%%%%%%%%%%%%%

In this paper, we proposed a {\em low complexity} precoding scheme based on the pairing of subchannels,
which achieves near optimal capacity for Gaussian MIMO channels with discrete inputs.
The low complexity feature relates to both the evaluation of the optimal precoder matrix
and the detection at the receiver. This makes the proposed scheme suitable for practical
applications, even when the channels are time varying and the precoder needs to be computed
for each channel realization.

The simple precoder structure, inspired by the X-Codes, enabled us to split
the precoder optimization problem into two simpler problems.
Firstly, for a given pairing and power allocation between pairs,
we need to find the optimal power fraction allocation
and rotation angle for each pair.
Given the solution to the first problem, the second problem is then
to find the optimal pairing and the power
allocation between pairs.

%At the lower level, the mutual information was optimized for each pair,
%then the optimal pairing was chosen and the power allocation to the %pairs was optimized.

For large $n$, typical of OFDM systems, we also discussed different heuristic
approaches for optimizing the pairing of subchannels.

The proposed precoder was shown to perform better than the Mercury/waterfilling strategy for both diagonal and non-diagonal MIMO channels.
Future work will focus on finding close to optimal pairings, and close to optimal power allocation strategies between pairs.

%%%%%%%%%%%%%%%%%%%%%%%%%%%%%%%%%%%%%%%%%%%%%%%%%%%%%%%%%%%%%%%%%%%%%%%%%%%%%%%%%%%%%%%%%%%%%%%%%%

% that's all folks
\end{document}